\documentclass[twoside]{article}
\usepackage{fleqn,espcrc2}
\usepackage{psfig}
\begin{document}
\title{On the $\Lambda$ to $\Sigma^0$ ratio from proton-proton
collisions\thanks{Supported in part by the Forschungszentrum J\"ulich and
the Australian Research Council.}}
\author{
A. Sibirtsev$^{1,2}$, K. Tsushima$^1$, W. Cassing$^{2}$, A. W.
Thomas$^1$ \\ \vspace{3mm}
{$^1$Special Research Center for the Subatomic Structure of Matter
and Department of Physics and Mathematical Physics,
University of Adelaide, SA 5005, Australia \\
$^2$ Institut f\"ur Theoretische Physik, Universit\"at Giessen,
D-35392 Giessen, Germany}}

\begin{abstract}
We compare the recent COSY data on the total $pp{\to}p\Lambda{K^+}$
and $pp{\to}p\Sigma^0{K^+}$ cross sections with the
predictions from two boson exchange models that either
are based solely on $\pi$ and $K$ exchange or include 
$\pi$, $\eta$ and $\rho$ exchange
as well as the virtual excitation of intermediate baryon
resonances. Both models are found to reproduce the data  
after the inclusion of final state hyperon-nucleon interactions.
Thus, within the experimental uncertainties, both models also roughly 
reproduce the strong dependence of the  $\Lambda$ to $\Sigma^0$ ratio
as a function of the excess energy near threshold, as well 
as at higher energies.
\end{abstract}
\maketitle

\vspace{6mm}

Recently the total $pp{\to}p\Lambda{K^+}$ and $pp{\to}p\Sigma^0{K^+}$
cross sections have been measured~\cite{Balewski,Bilger,Sewerin} at
the COoler  SYnchrotron (COSY) at invariant collision energies
$\sqrt{s}$ near the production threshold. It was
observed~\cite{Sewerin} that the ratio, $R$, of the cross sections
for $\Lambda$ and $\Sigma$ production at the same excess energies, 
$\epsilon{=}\sqrt{s}{-}m_p{-}m_Y{-}m_K{<}13$~MeV, varies from about
20 -- 30 at threshold to $R{\simeq}$2.5 at high energies, 
$\epsilon{\ge}1$~GeV~\cite{LB}. As a possible explanation it has
been suggested that this large ratio close to threshold might 
arise from a strong $\Sigma{\to}\Lambda$ conversion by the
final state interaction.

Very recently the $\Lambda$ to $\Sigma^0$ ratio from
$pp$ collisions was studied in Ref.~\cite{Juelich1}.
It was found~\cite{Juelich1} that the experimentally observed 
suppression of the ratio of $\Sigma^0$ to $\Lambda$ hyperon
production may be explained, at least in the region up to 
15 MeV above threshold, by a destructive
interference between the pion and kaon exchange.

To make a further investigation of these remarkable features~\cite{Sewerin}
we consider here two types of
boson exchange models~\cite{Sibirtsev1,SibCa,Tsushima1}, that
have both been used quite often~\cite{OBEM}
in the analysis of strangeness production in nucleon-nucleon interactions.
To get more insight into the dynamics we compare the model calculations
simultaneously with the $pp{\to}p\Lambda{K^+}$ and
$pp{\to}p\Sigma^0{K^+}$ cross
sections  as a function of the excess energy, $\epsilon$. Furthermore,
to obtain a more decisive conclusion we perform
the calculations with the model parameters as they were fixed
some years ago in Refs.~\cite{Sibirtsev1,SibCa,Tsushima1}, prior to
the COSY data in Refs.~\cite{Balewski,Bilger,Sewerin}, rather 
than adjusting these parameters again. This also provides a crucial 
test of the predictive power of these models.

\begin{figure}[t]
\phantom{aa}\vspace{-1mm}\hspace{-10mm}
\psfig{file=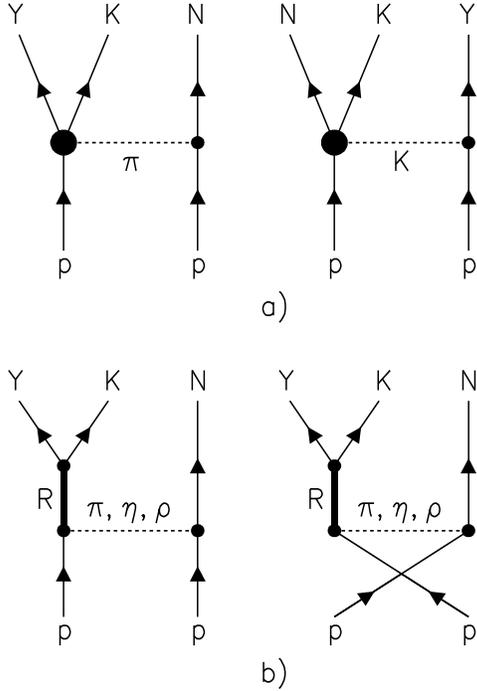,height=10cm,width=9cm}
\vspace{-14mm}
\caption[]{The diagrams for $pp{\to}NYK$ reactions 
in the $\pi{+}K$-exchange model (a) 
and within the resonance model (b).}
\label{kilian7}
\vspace{-3mm}
\end{figure}

We recall that the $\pi{+}K$-exchange model leads to strangeness 
production through $\pi$ and
$K$-meson exchanges, with the ${\pi}N{\to}YK$ and $KN{\to}KN$
scattering amplitudes (cf. Fig.~\ref{kilian7}a) evaluated from 
the data~\cite{LB}. The coupling
constant and the form factor (FF) at the $NN\pi$ vertex were
taken from the Bonn-J\"ulich model~\cite{Juelich}, while the
parameters at the $NYK$ vertex were fitted~\cite{Sibirtsev1} to
$pp{\to}NYK$ data at $\epsilon{\ge}1$~GeV.
Because of the lack of data, the interference between the $\pi$
and $K$-meson could not be unambiguously fixed and for
simplicity was neglected.
For further details of the model and its explicit expressions we
refer the reader to Ref.~\cite{SibCa}.

\begin{figure}[h]
\phantom{aa}\vspace{-15mm}\hspace{-3mm}
\psfig{file=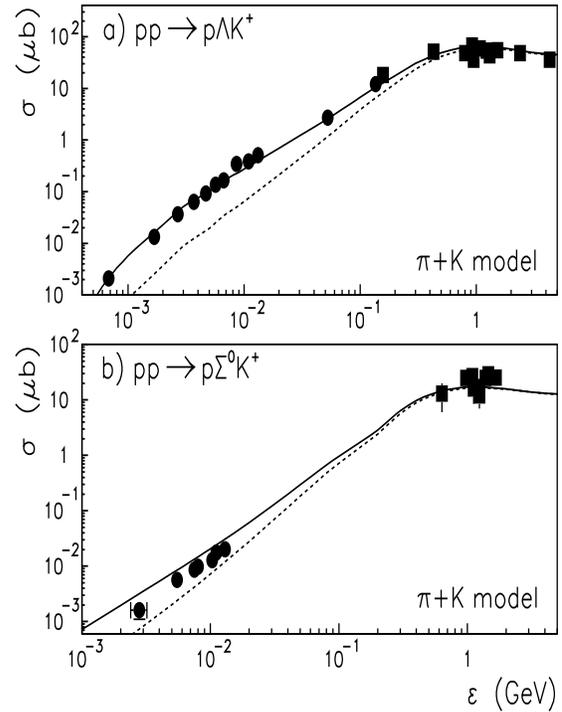,height=11cm,width=7.9cm}
\vspace{-14mm}
\caption[]{The  $pp{\to}p\Lambda{K^+}$ (a) and
$pp{\to}p\Sigma^0{K^+}$ (b) cross sections as a function
of the excess energy $\epsilon$. The circles show the
data from COSY~\protect\cite{Balewski,Bilger,Sewerin}
while the squares are from Ref.~\protect\cite{LB}. The lines
show the calculations within the $\pi{+}K$-exchange model with (solid) and
without (dashed) FSI.}
\label{kilian3}
\vspace{-2mm}
\end{figure}

The predictions of this model
for the $pp{\to}p\Lambda{K^+}$ and $pp{\to}p\Sigma^0{K^+}$ cross
sections are shown  by the dashed lines in Fig.~\ref{kilian3} 
together with the data from Ref.~\cite{LB}(squares) and the recent data
(circles) from COSY~\cite{Balewski,Bilger,Sewerin}. The
calculations describe the experimental results reasonably well at
high energies but substantially underestimate the data
at $\epsilon{\le}$100~MeV, because 
the final state interaction (FSI) between the proton and
the hyperon has been neglected~\cite{FSI}.

Among the different semi-phenomenological corrections to
the FSI~\cite{FSI} we employ here the Jost approximation~\cite{FSI1,FSI2}
and convolute the production amplitude with the function
\begin{equation}
J_s{=}\frac{q{-}i\alpha}{q{+}i\beta}, \hspace{1mm} {\rm where}
\hspace{2mm} \alpha{+}\beta{=}\frac{2}{r_s} \hspace{1mm} {\rm and}
\hspace{2mm} \alpha{\times}\beta{=}\frac{2}{r_s\,a_s},
\label{cor1}
\end{equation}
where $a_s$ and $r_s$ are the scattering length and the effective
range for the S-wave hyperon-proton interaction, respectively,
which were taken for the singlet and triplet states from
Ref.~\cite{Range}. The function~(\ref{cor1})
is almost identical~\cite{FSI2} to the FSI correction used in
the experimental analysis of Refs.~\cite{Balewski,Sewerin} at low
$\epsilon$, where the FSI dominates. The solid lines in
Fig.~\ref{kilian3} show the $\pi{+}K$-model calculations
including FSI. Clearly the calculation describes the $p K^+ \Lambda$
channel remarkably well, while the $p K^+ \Sigma^0$ cross section is 
slightly overestimated very close to threshold.
This might also be attributed to an uncertainty in the $\Sigma^0
p$ s-wave scattering parameters, $a_s$ and $r_s$.

On the other hand, the resonance model~\cite{Tsushima1} 
generates strangeness
production by  $\pi$, $\eta$ and $\rho$-meson
exchange with the excitation of intermediate baryon
resonances, $R$ ($N(1650)$, $N(1710)$, $N(1710)$ and ${\Delta}(1920)$) 
(cf. Fig. \ref{kilian7}b), 
which can couple to the $\Lambda{K}$ and $\Sigma{K}$ states~\cite{PDG}.

The parameters of the $RYK$ and $RN\pi$
vertices in Fig. 1b were fixed~\cite{Tsushima2} using the 
${\pi}N{\to}\Lambda{K}$ and  ${\pi}N{\to}\Sigma{K}$
data from Ref.~\cite{LB}, which were available prior to the COSY data. 
The couplings and formfactors at the $NN\pi$, $NN\eta$,
$NN\rho$, $RN\eta$ and $RN\rho$ vertices were either taken from 
Ref.~\cite{Juelich} or fitted~\cite{Tsushima1}  to the
$pp{\to}NYK$ data at high energies, $\epsilon{\ge}1$~GeV.
Note again, that the baryonic resonance couplings are well
controlled by the presently available
resonance properties~\cite{PDG}.

\begin{figure}[h]
\phantom{aa}\vspace{-12mm}\hspace{-3mm}
\psfig{file=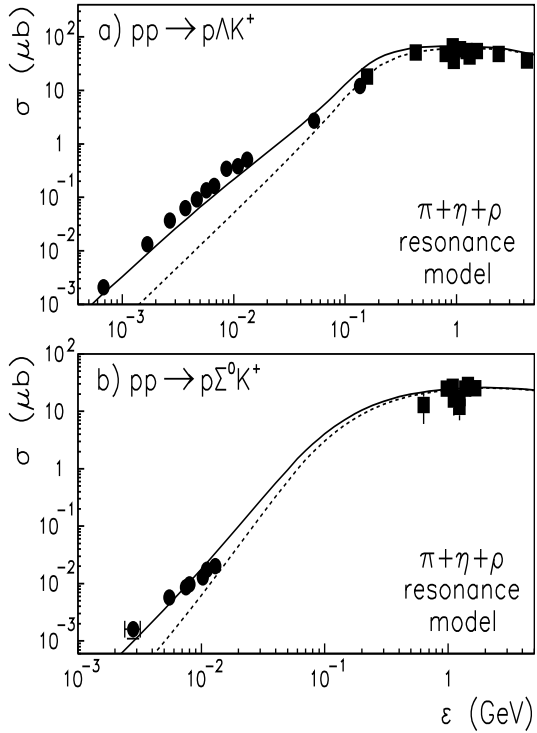,height=11cm,width=8.2cm}
\vspace{-14mm}
\caption[]{The  $pp{\to}p\Lambda{K^+}$ (a) and
$pp{\to}p\Sigma^0{K^+}$ (b) cross sections as a function
of the excess energy $\epsilon$. The circles show the
data from COSY~\protect\cite{Balewski,Bilger,Sewerin}
while the squares are from Ref.~\protect\cite{LB}. The lines
show the calculations within the  resonance model with (solid) and
without (dashed) FSI.}
\label{kilian4}
\vspace{-3mm}
\end{figure}

The resonance model calculations for the $\Lambda$ and $\Sigma^0$ channel
without FSI are shown by
the dashed lines in Fig.~\ref{kilian4}, while the solid lines
 show the resonance model results obtained with the same FSI~(\ref{cor1})
as in Fig. \ref{kilian3}.

Keeping in mind that the parameters for both
versions of the boson exchange models were not adjusted to the recent
data~\footnote{The parameters
of the $\pi{+}K$-exchange model were fixed in 1995~\cite{Sibirtsev1},
while the resonance model parameters were fixed
in 1997~\cite{Tsushima1} -- i.e., before the appearance of any near
threshold COSY data~\cite{Balewski,Bilger,Sewerin}.} we conclude that
the agreement between the model
predictions~\cite{Sibirtsev1,SibCa,Tsushima1} and
the experimental results~\cite{Balewski,Bilger,Sewerin}
is quite reasonable. It is important to note that both
calculations reproduce simultaneously the data
at low and high energies.  Since both models were  
originally fixed at high energy, one might conclude that 
the cross sections at low energy are a strict
consequence of the underlying reaction mechanism.

\begin{figure}[h]
\phantom{aa}\vspace{-11mm}\hspace{-3mm}
\psfig{file=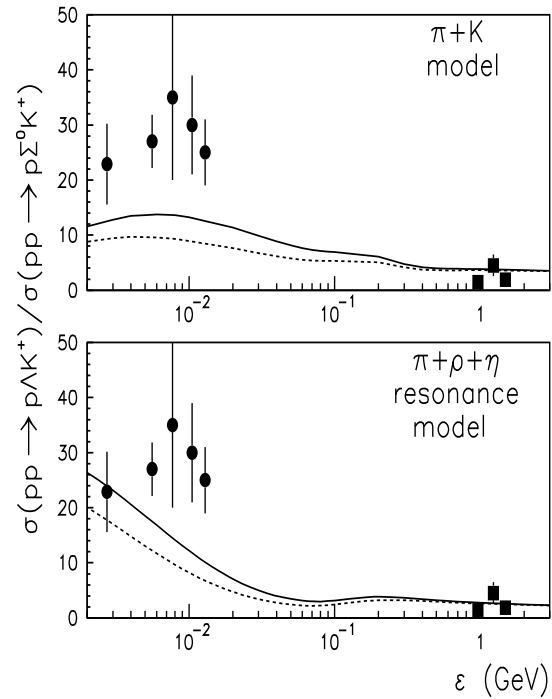,height=10.7cm,width=8.2cm}
\vspace{-17mm}
\caption[]{The  ratio of the $pp{\to}p\Lambda{K^+}$  and
$pp{\to}p\Sigma^0{K^+}$  cross sections as a function
of the excess energy $\epsilon$. The circles show the
data from COSY~\protect\cite{Sewerin}
while the squares are from Ref.~\protect\cite{LB}. The lines
show the results  with (solid) and
without (dashed) FSI calculated by the $\pi{+}K$-exchange (upper part)
and the  resonance model (lower part).}
\label{kilian1}
\vspace{-3mm}
\end{figure}

Finally, the data~\cite{Sewerin} on the $\epsilon$-dependence of
the ratio of $pp{\to}p\Lambda{K^+}$ to $pp{\to}p\Sigma^0{K^+}$
cross sections are shown in Fig.~\ref{kilian1} together with the
results from the $\pi{+}K$-exchange  (upper part) and the resonance 
model (lower part). The
dashed lines show the results obtained without FSI while the
solid lines indicate the calculations with FSI. The
comparison in Fig.~\ref{kilian1} is given using a linear scale for $R$ 
and within about two experimental errors bars the
agreement between data and calculations is reasonable.
Apparently, by adjusting the
coupling constants and cut-off parameters of the formfactors one can
obtain a much better description of the
data~\cite{Balewski,Bilger,Sewerin} which, 
however, is not the aim of our present
investigation.

Our results indicate that the
strong variation of the ratio $R$ with the excess energy
$\epsilon$ should not be considered as an extraordinary feature
of the strangeness production process since the
two different approaches, fixed some years ago at high energy,
are obviously able to reproduce the data to better than a factor of 2.
This finding also shows that the
present data on the total cross sections are not sensitive to
the details of the model and that it is still not
possible to fix the explicit contribution from the kaon
exchange channel, which is of fundamental interest.

We point out that a fully  self-consistent model should include
all available $\pi$, $\eta$, $\sigma$, $\rho$, $K$
and possibly $K^\ast$ and higher mass mesonic exchanges as well
as the excitation of all available intermediate baryonic resonances
and the direct kaon emission processes. The large number of 
parameters involved in such a model might be fixed using the $NN{\to}NYK$
data. However, this would require a larger number of experimental points 
than presently available~\cite{Balewski,Bilger,Sewerin,LB}.

In spite of these comments, the theoretical progress resulting from the new
COSY data~\cite{Balewski,Bilger,Sewerin,LB} is significant.
The role of FSI, as well
as a direct experimental evaluation of the hyperon-nucleon interaction
parameters from the data~\cite{Balewski2}, can now be well
understood phenomenologically. This is an important step for 
{}further theoretical and experimental analysis of
differential cross section data on strangeness production in 
nucleon-nucleon collisions.

More detailed constraints on the theory can be obtained from
differential cross section data at higher excess energies. 
It is crucial that the contribution from $K$-meson exchange may be 
directly investigated from polarization measurements as 
argued in Ref.~\cite{DISTO}, or by a partial wave decomposition of the
Dalitz representation as advocated in Ref.~\cite{Tsushima1}. 
The separation of the kaon
exchange contribution would provide information on the 
$NYK$ coupling constants, which could be compared with 
the $SU(N)$ symmetry predictions
and other experimental determinations~\cite{Sibirtsev1,Sibirtsev2}.

The authors like to acknowledge  discussions with W. Eyrich and
K. Kilian that stimulated this study.
A.S. would also like to acknowledge the warm hospitality at the CSSM during
his visit. This work was supported by the
Australian Research Council and the Forschungszentrum J\"ulich.

\end{document}